\documentclass[%
 aip,%
 jmp,%
 amsmath,amssymb,%
 reprint,%
]{revtex4-2}

\usepackage{graphicx}
\usepackage{dcolumn}
\usepackage{bm}
\usepackage{hyperref}

\begin{document}

\title{Hybrid CO\textsubscript{2}-Ti:sapphire laser with tunable pulse duration\\ for mid-infrared-pump terahertz-probe spectroscopy}

\author{Matthias Budden}
\thanks{Co-first authors with equal contribution}
\affiliation{Max Planck Institute for the Structure and Dynamics of Matter, 22761 Hamburg, Germany}
\author{Thomas Gebert}
\thanks{Co-first authors with equal contribution}
\affiliation{Max Planck Institute for the Structure and Dynamics of Matter, 22761 Hamburg, Germany}
\author{Andrea Cavalleri}
\email{andrea.cavalleri@mpsd.mpg.de}
\affiliation{Max Planck Institute for the Structure and Dynamics of Matter, 22761 Hamburg, Germany}
\affiliation{Department of Physics, Clarendon Laboratory, University of Oxford, Oxford OX1 3PU, United Kingdom}

\date{\today}

\begin{abstract}
Ultrafast optical excitation with intense mid-infrared and terahertz pulses has emerged as a new tool to control materials dynamically. As most experiments are performed with femtosecond pulse excitation, typical lifetimes of most light-induced phenomena in solids are of only few picoseconds. Yet, many scientific applications require longer drive pulses and lifetimes. Here, we describe a mid-infrared pump – terahertz-probe setup based on a CO\textsubscript{2} laser seeded with 10.6\,$\mathrm{\mu m}$ wavelength pulses from an optical parametric amplifier, itself pumped by a Ti:$\mathrm{Al_2O_3}$ laser. The output of the seeded CO\textsubscript{2} laser produces high power pulses of nanosecond duration, which are synchronized to the femtosecond laser. Hence, these pulses can be tuned in pulse duration by slicing their front and back edges with semiconductor-plasma mirrors irradiated by replicas of the femtosecond seed laser pulses. Variable pulse lengths from 5\,ps to 1.3\,ns are achieved, and used in mid-infrared pump, terahertz-probe experiments with probe pulses generated and electro-optically sampled by the femtosecond laser.
\end{abstract}

\maketitle

\section*{Introduction}
Intense mid-infrared (MIR) and terahertz light pulses are of strong interest for nonlinear ultrafast spectroscopy as well as for coherent manipulation of materials properties. Recently, such pulses have enabled the discovery of unconventional phenomena such as the formation of magnetic polarization in antiferromagnets~\cite{disaPolarizingAntiferromagnetOptical2020}, ferroelectricity in paraelectrics~\cite{liTerahertzFieldInduced2019, novaMetastableFerroelectricityOptically2019}, new topological phases~\cite{wangObservationFloquetBlochStates2013}, as well as non-equilibrium high $\mathrm{T_c}$ superconductivity in cuprates~\cite{faustiLightInducedSuperconductivityStripeOrdered2011, huOpticallyEnhancedCoherent2014} and molecular organic materials \cite{mitranoPossibleLightinducedSuperconductivity2016, cantaluppiPressureTuningLightinduced2018, buzziPhotomolecularHighTemperatureSuperconductivity2020, buddenEvidenceMetastablePhotoinduced2020}. Typically, these experiments are performed using sub-picosecond pulses with high peak electric field ($\sim$\,5 – 10\,MV/cm). The lifetime of the induced transient state is therefore often limited to a few picoseconds. In the pursuit of longer-lived, functionally relevant photo-induced states, there is an increasing demand for laser sources producing mid-infrared and terahertz pulses with tunable duration from hundreds of femtoseconds to hundreds of picoseconds~\cite{buddenEvidenceMetastablePhotoinduced2020}.

Laser sources based on narrow band gap semiconductors, such as IV-VI compounds (lead salts) and wavelength tunable quantum cascade lasers readily produce radiation in the mid-infrared region of the electromagnetic spectrum ($\sim$\,2.5 – 30\,$\mathrm{\mu m}$ wavelength). However, these sources often require cryogenic cooling~\cite{tackeLeadSaltLasers2001}, and their emitted power is  not sufficient to reach sufficiently intense pulses, as field strengths in excess of few MV/cm are required for nonlinear excitation of materials.\\
Typically, high peak power mid-infrared pulses are generated via optical parametric amplification and frequency mixing starting from ultra short light pulses generated by conventional solid state lasers (e.g.\ Ti:$\mathrm{Al_2O_3}$ or Yb:YAG). This approach yields pulses with energies of 10 - 50\,$\mathrm{\mu J}$ and allows to reach MV/cm peak electric fields only for pulses with sub-picosecond durations. Although gas lasers based on CO ($\mathrm{\lambda}$ = 5 - 6\,$\mathrm{\mu m}$) and CO\textsubscript{2} ($\mathrm{\lambda}$ = 9 - 11\,$\mathrm{\mu m}$) can easily produce high energy mid-infrared pulses~\cite{endoGasLasers2007, mesiatsPulsedPower2005, beaulieuTransverselyExcitedAtmospheric1970}, they are typically not mode-locked resulting in pulse-to-pulse time and intensity fluctuations that make pump-probe experiments with picosecond precision difficult.

\begin{figure}[h!]
\includegraphics[width=9cm]{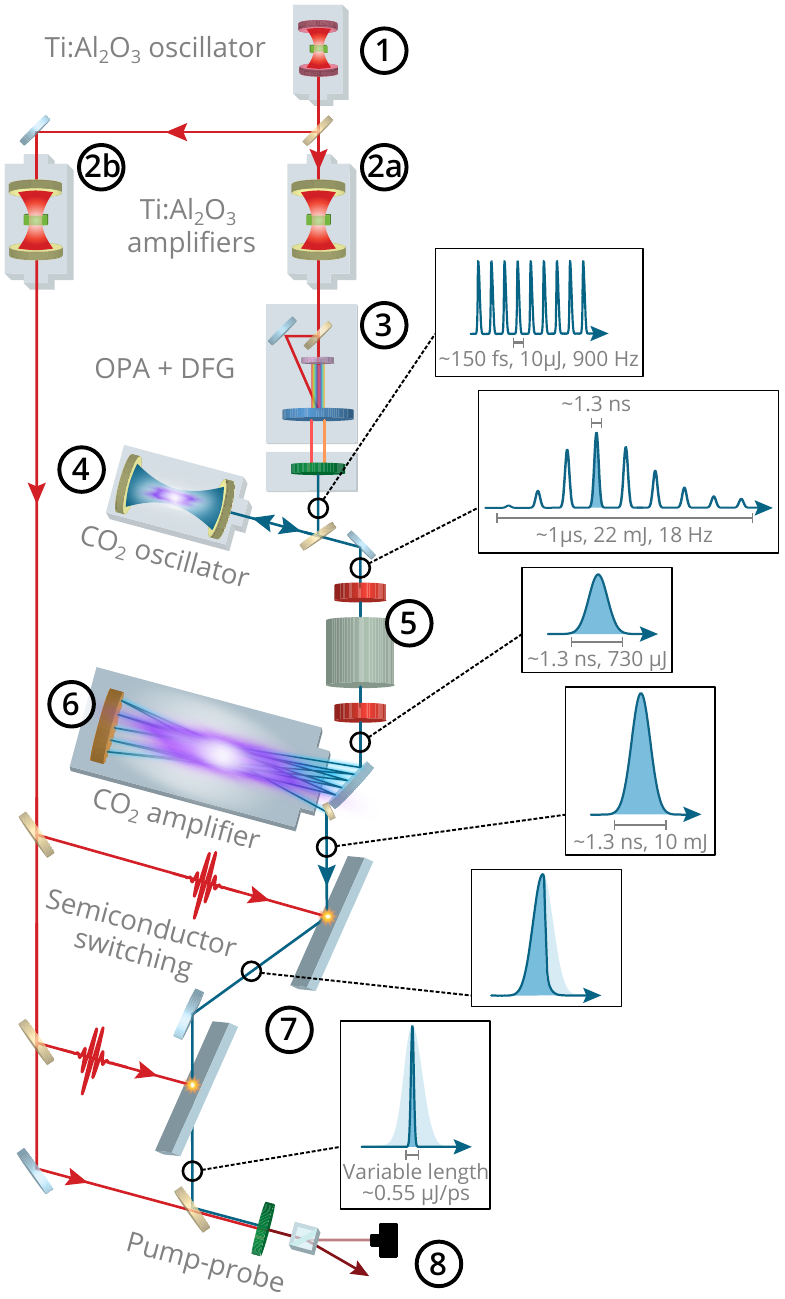}
\caption{\label{fig_1}Experimental setup for the generation of high power, pulse length tunable mid-infrared pulses
(1) Ti:$\mathrm{Al_2O_3}$ oscillator (f = 80\,MHz, $\lambda$ = 800\,nm, E = 6.25\,nJ), (2) Ti:$\mathrm{Al_2O_3}$ amplifiers (a: E = 5\,mJ, b: E = 7\,mJ), (3) OPA and DFG ($\mathrm{\lambda}$ = 10.6\,$\mathrm{\mu m}$) (4) Injection seeded CO\textsubscript{2} oscillator ($\mathrm{\lambda}$ = 10.6\,$\mathrm{\mu m}$) (5) CdTe Pockels cell selects most intense pulse (6) CO\textsubscript{2} multipass amplifier (7) Semiconductor switching (8) Cross-correlation and pump-probe experiment.}
\end{figure}
Here we present a hybrid optical source based on a combination of ultrafast Ti:$\mathrm{Al_2O_3}$ and CO\textsubscript{2} lasers to perform pump-probe measurements using intense MIR pulses with tunable duration. Figure~\ref{fig_1} displays a schematic of the experimental setup. A CO\textsubscript{2} laser chain is seeded with radiation generated by an ultrafast optical parametric amplifier to generate mid-infrared pulses at 10.6\,$\mathrm{\mu m}$ that are optically synchronized to the Ti:$\mathrm{Al_2O_3}$ source, enabling experiments with sub-picosecond time resolution.

\section*{Generation of mode-locked mid-infrared pulses with CO\textsubscript{2} lasers}
\begin{figure}[b!]
\includegraphics[width=\linewidth]{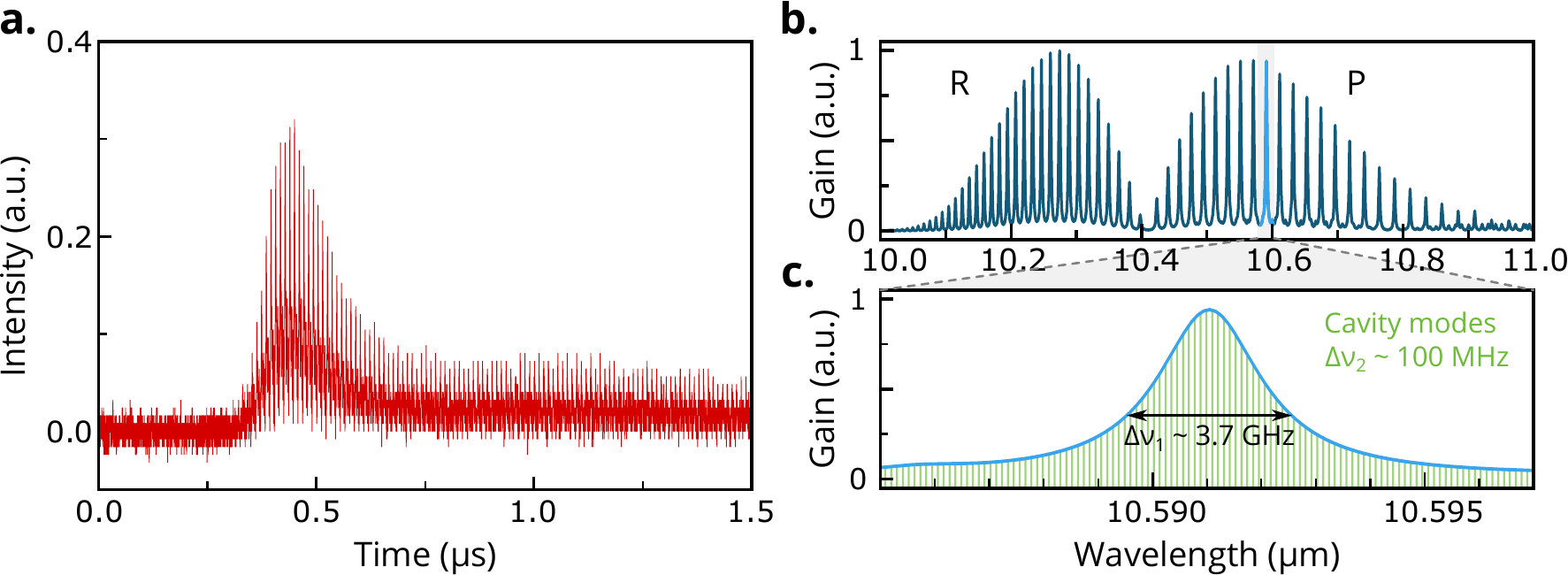}
\caption{\label{fig_2}Amplified spontaneous emission and multimode operation of the CO\textsubscript{2} laser oscillator. \textbf{a.} Time-intensity profile of the pulse emitted from a commercial CO\textsubscript{2} laser showing the superposition of several longitudinal modes. The most intense mode causes distinct spikes to appear every $\sim$\,10\,ns corresponding to the roundtrip time of the $\sim$\,1.5\,m cavity. The pulse has an envelope of a few microseconds. \textbf{b.} Calculated gain spectrum of molecular CO\textsubscript{2} at ambient pressure. Rotational transitions with a spacing of $\sim$\,55 GHz form the R- and P-branches. A grating reflector within the cavity selects only one transition around 10.6\,$\mathrm{\mu m}$. \textbf{c}. Calculated gain spectrum of the selected rotational line compared to the frequency spacing of the longitudinal cavity modes. The $\sim$\,3.7\,GHz gain bandwidth is much wider than the 100\,MHz spacing of the longitudinal cavity modes resulting in arbitrary excitation of multiple cavity modes when the laser is amplifying spontaneous emission.}
\end{figure}

\begin{figure}[h!]
\includegraphics[width=\linewidth]{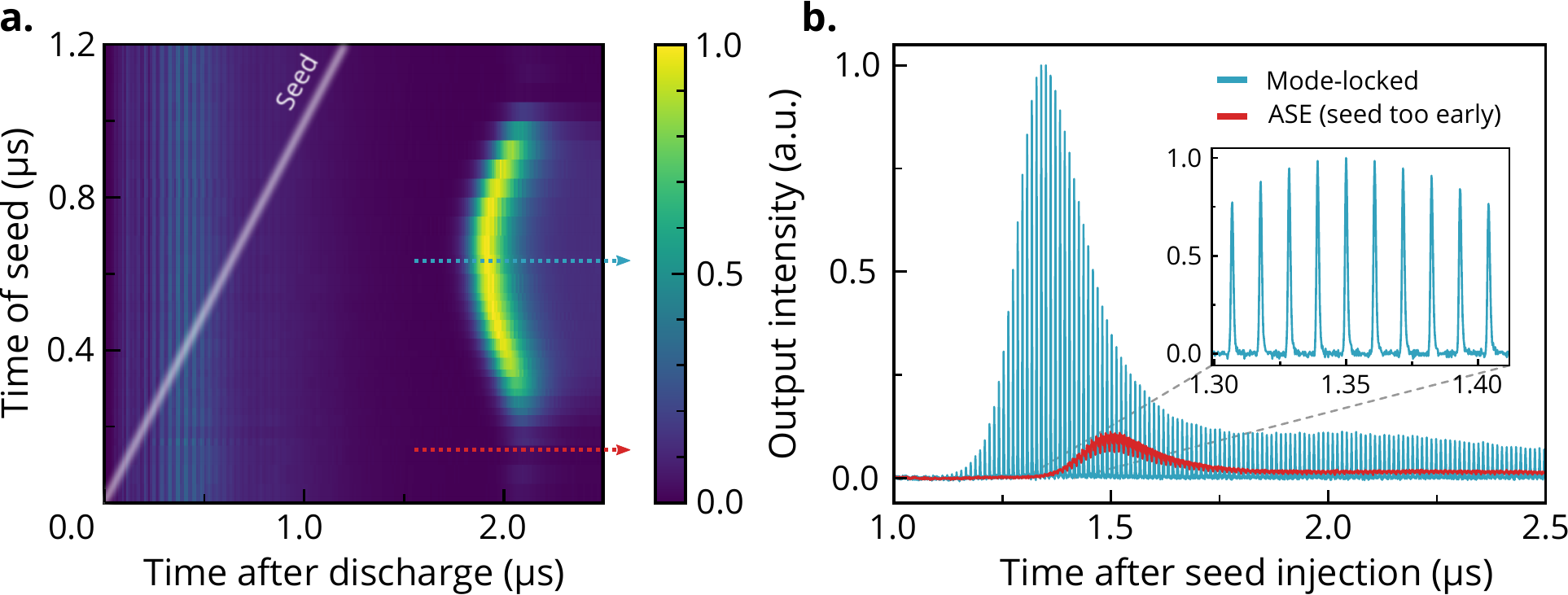}
\caption{\label{fig_3}CO\textsubscript{2} laser injection seeding
\textbf{a.} Map of the measured CO\textsubscript{2} laser output traces vs. time of the injected seed pulse. The color indicates the output intensity. Both time axes are referenced to the trigger of the plasma discharge. The discharge leads to a stable electronic noise floor within the first microsecond after the trigger. The seed injection time is shown as a white line. The two dashed lines correspond to single time traces shown in b. \textbf{b.} Averaged output intensity time traces of the CO\textsubscript{2} laser are shown for the case of amplified spontaneous emission when the 100\,fs long seed pulse is injected too early (red curve) as well as for optimal seeding (blue curve). The inset shows a section of a single-shot time trace of the seeded laser indicating perfect mode-locking.}
\end{figure}

The gain spectrum of the CO\textsubscript{2} laser consists of multiple lines with a bandwidth of about 3.7\,GHz and about 55\,GHz spacing originating from the vibrational-rotational transitions in the laser gas (see Fig.~\ref{fig_2}b). This bandwidth is much larger than the typical spacing of cavity modes of about 100\,MHz (see Fig.~\ref{fig_2}c) so that multiple longitudinal modes can be excited at the same time even if the gain is limited to a single rotational line (e.g.\ with a grating reflector). Figure~\ref{fig_2}a displays the typical single shot output of a commercial transversely excited CO\textsubscript{2} laser, revealing multi-mode operation arising from amplified spontaneous emission.\\
Whilst mode-locking techniques based on saturable absorbers, electro- or acousto-optic modulators facilitate the formation of reproducible high power output pulses \cite{gibsonGenerationBandwidthlimitedPulses1974, nurmikkoSinglemodeOperationMode1971, woodPassiveQswitchingCeCO21967}, they are not well suited to synchronize the generated MIR pulses to a probe beam. This limitation can be overcome by externally seeding the CO\textsubscript{2} laser cavity, both as a means of active mode locking and to facilitate synchronization \cite{bourdetActiveModeLocking1987, bridgesSpontaneousSelfpulsingCavity1969}. For this purpose, we generated 150\,fs long 10.6\,$\mathrm{\mu m}$ wavelength pulses in a 1.5\,mm thick GaSe crystal by difference frequency mixing of the signal (1490\,nm) and idler (1730\,nm) outputs of a home built optical parametric amplifier (OPA). The OPA was pumped with 3\,mJ pulse energy, 100\,fs long pulses from a commercial Ti:$\mathrm{Al_2O_3}$ regenerative amplifier (see Fig.~\ref{fig_1}). A fraction of these femtosecond pulses ($\sim$\,60 pJ) was injected through the semi-transparent (R: 80\,\%, T: 20\,\%) front window into the cavity of the CO\textsubscript{2} laser. The bandwidth of the injected femtosecond pulses was spectrally filtered to match the narrower gain bandwidth of the CO\textsubscript{2} laser cavity (see Fig.~\ref{fig_2}b, c) by spatial selection of the first diffraction order of the intracavity grating reflector, which acted as a rear mirror only for the desired rotational line around 10.6\,$\mathrm{\mu m}$. The injection of these seed pulses induces an active mode locking resulting in a train of output pulses, synchronized to the femtosecond seed laser \cite{babzienSolidstateSeedingHigh2016}. A typical time-intensity trace of the seeded output of the CO\textsubscript{2} oscillator is shown in Fig.~\ref{fig_3}b (blue curve). The peak intensity of the pulse train was about one order of magnitude higher than in the non-seeded case (red curve) due to the phase locking of all longitudinal cavity modes induced by the seed. The single pulses in the train were $\sim$\,1.3\,ns long, limited by the available gain bandwidth, and showed intensity fluctuations of less than 1\,\%. These residual fluctuations can be attributed to small temporal variations of the electrically triggered plasma build-up, which translated into a time jitter of the gain curve. For the specific laser used in this work, the optimal timing for seed injection was found to be $\sim$\,700 ns after the plasma excitation is triggered. Under these conditions, the pulse build-up time was reduced by $\sim$\,150\,ns as compared to unseeded operation and the output intensity was highest.\\
The most intense pulse in the train was selected by a custom-built pulse picker based on the combination of a CdTe Pockels-cell (Fig.~\ref{fig_1}-5) and a wire-grid polarizer. To reach the required pulse energy for the envisioned experiments, this isolated pulse was further amplified from $\sim$\,0.7\,mJ to $\sim$\,10\,mJ in a subsequent CO\textsubscript{2} laser amplifier, built from a commercial laser that we modified to operate as a 10-pass amplifier (Fig.~\ref{fig_1}-6) at a repetition rate of 18\,Hz.

\section*{Mid-infrared pulse duration control}
Although a pressure or electric field (through the AC Stark effect) broadened spectral gain can support the amplification of picosecond CO\textsubscript{2} laser pulses, it requires either an amplifier with a gas cell pressurized at 10 – 15\,bar \cite{patelCollisionBroadeningHigh1973, alcockGenerationDetection150psec1974, corkumAmplificationPicosecond101985} or multi-stage preamplification for the generation of laser fields with intensities of more than 5\,GW/cm² \cite{haberbergerFifteenTerawattPicosecond2010a}. As the mid-infrared pulses produced in our setup are inherently synchronized to a femtosecond near infrared source, another viable option for tailoring their pulse duration is the use of electron-hole plasmas in photoexcited semiconductors as ultrafast shutters. This technique, first described in detail in Ref.~\cite{alcockFastScalableSwitching1975}, has been applied in different laser systems for external pulse shortening~\cite{corkumAmplificationPicosecond101985, jamisonGenerationPicosecondPulses1978} as well as for the reflection of a single pulse out of a laser cavity~\cite{apollonovSelectionHighpowerNanosecond1979, debekkerGenerationVeryShort1990, marchettiControlledDumpingPulses2001}.

\begin{figure}[h!]
\includegraphics[width=\linewidth]{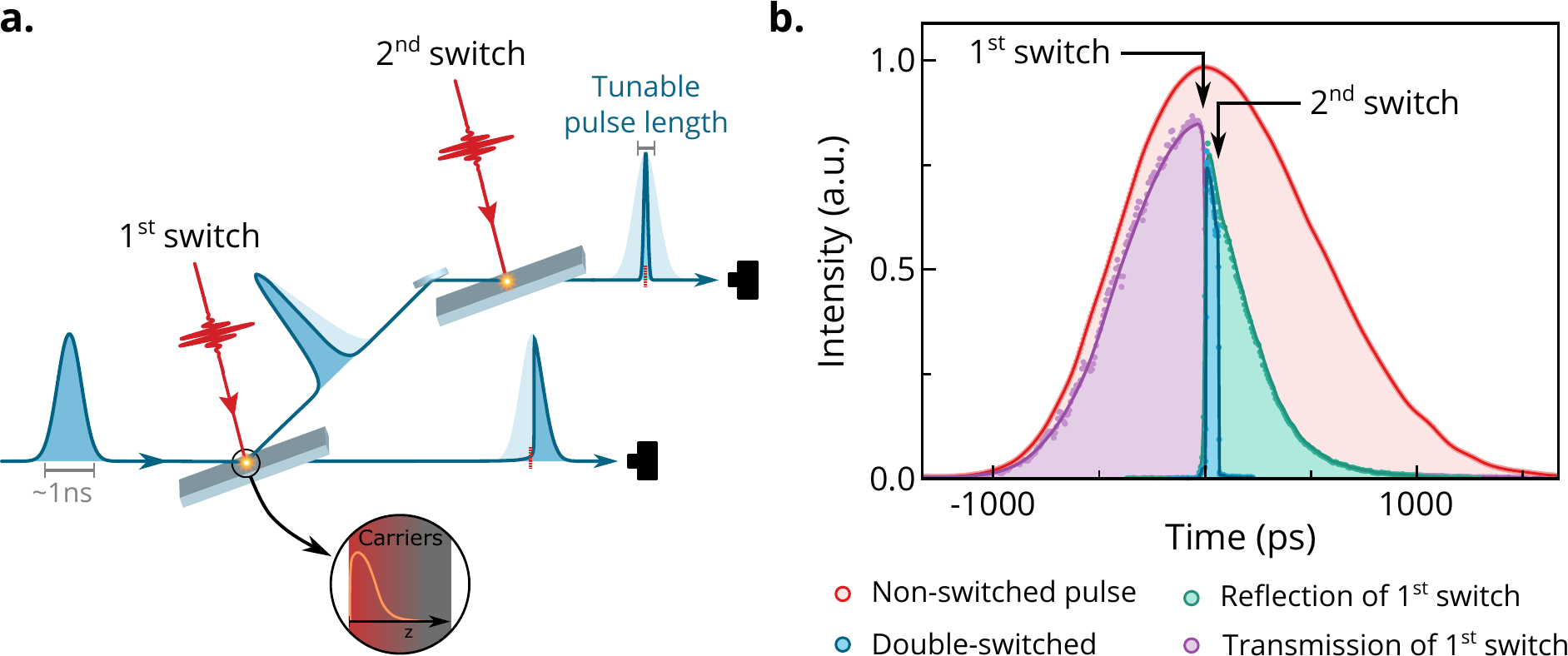}
\caption{\label{fig_4}Principle of semiconductor switching. \textbf{a.} Schematic of the setup for generation of continuously adjustable pulse lengths. The incoming MIR CO\textsubscript{2} laser pulse (blue) impinges on a semiconductor at Brewster's angle and its leading edge is fully transmitted. A second short NIR pulse (800\,nm wavelength, $\sim$\,100\,fs duration, red) excites an electron-hole plasma at spatio-temporal overlap with the MIR pulse. Within few picoseconds, the carrier density rises, which renders the semiconductor highly reflective and slices the MIR pulse. A subsequent switch can cut the trailing edge so that precise pulse length tunability is possible by adjusting the optical delay between the two NIR slicing pulses. \textbf{b.} Cross correlation curves of the unswitched (red) CO\textsubscript{2} laser beam as well as for single transmissive (violet), single reflective (green) and double switched pulses (blue). Upon switching, the semiconductor reflectivity changes within less than 5\,ps creating sharp edges in the pulse.
}
\end{figure}
In our implementation, shown schematically in Fig.~\ref{fig_4}a, the p-polarized mid-infrared laser pulse was transmitted through an optically flat semiconducting plate set at Brewster’s angle to suppress any reflection in the unpumped state. A short near-infrared laser pulse, derived from the Ti:$\mathrm{Al_2O_3}$ chain used to seed the CO\textsubscript{2} oscillator was used to excite a plasma of electron-hole pairs. Optical excitation made the semiconductor surface highly reflective for all frequencies lower than the electron-hole-density-dependent plasma frequency. After optical excitation, the critical carrier density for full reflection can be obtained from:
\begin{equation}
n_c^* = \frac{\epsilon_0\epsilon_r m^* \omega_l^2}{e^2},
\end{equation}
where $\omega_l$ is the laser frequency and $\epsilon_0$, $\epsilon_r$, $m^*$, and $e$ denote the vacuum and semiconductor dielectric constants, the effective carrier mass and the electron charge, respectively. To create a reflective surface for CO\textsubscript{2} laser radiation at 10.6\,$\mathrm{\mu m}$ wavelength, a carrier density in excess of $\mathrm{10^{18} - 10^{19}\,cm^{-3}}$ is needed depending on the type of semiconductor. Si, Ge, CdTe, and GaAs all feature high mid-infrared transparency and critical carrier densities that can be achieved with modest ($<$\,10\,mJ/cm²) excitation fluences.

The combination of a reflection and a transmission switching enabled the generation of short pulses of variable length that could be adjusted by appropriately delaying the control pulses. In order to compensate the optical path in the CO\textsubscript{2} lasers of $\sim$\,400\,m, the near-infrared excitation pulse for semiconductor switching was derived from a second Ti:$\mathrm{Al_2O_3}$ amplifier seeded with a later pulse from the master oscillator (see Fig.~\ref{fig_1}). The remaining optical delay of $\sim$\,2\,m was compensated with an optical delay line to achieve accurate timing.

To perform pump-probe experiments using these pulses we required high switching efficiency, wide pulse length tuning range (10\,ps to more than 500\,ps), and a high background suppression of at least 1:1000. To achieve a high switching contrast, good mid-infrared transparency in the unexcited state and high reflectivity in the photoexcited state were required.

\begin{figure}
\includegraphics[width=\linewidth]{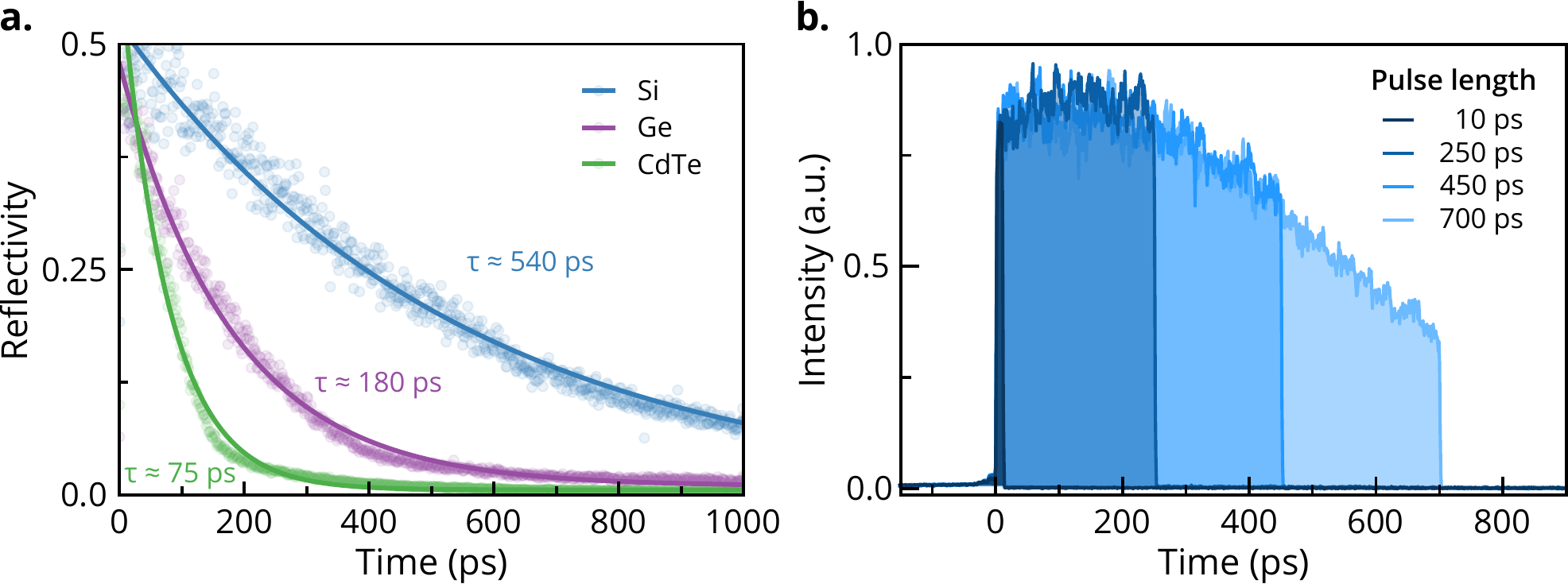}
\caption{\label{fig_5}Time profile of semiconductor plasma reflectivity and switched pulses
\textbf{a.} The temporal evolution of the reflected fraction of the incoming pulse is shown for three different slicing materials. Note that the power of the 800\,nm control beam was adjusted for every switch for maximum efficiency without damaging the material. \textbf{b.} Time profile of the intensity of the 10.6\,$\mathrm{\mu m}$ pulses after slicing on two silicon wafers in reflection and transmission. The traces are a result of a cross-correlation measurement following the procedure described in the text.}
\end{figure}

Figure~\ref{fig_4}b displays typical time-traces of the resulting 10.6\,$\mathrm{\mu m}$ pulses before and after switching using silicon both as a reflection and transmission switch. These measurements were performed by cross-correlation of the mid-infrared pulse with an ultrafast NIR pulse (100\,fs pulse length, $\mathrm{\lambda = 800\,nm}$) via sum-frequency generation in a 2\,mm thick GaSe crystal. Different semiconductors feature distinct decay dynamics~\cite{rollandGeneration130fsecMidinfrared1986, elezzabiGeneration1psInfrared1994, elezzabiUltrafastSwitchingCeCO21995} of the electron-hole plasma. Figure~\ref{fig_5}a shows a comparison of the time-dependent 10.6\,$\mathrm{\mu m}$ reflectivity of a single semiconductor switch made of silicon, germanium and cadmium telluride. These curves were extracted from cross-correlation measurements shown exemplarily in Fig.~\ref{fig_4}b by calculating the time dependent ratio of the reflected intensity from the semiconductor to the intensity of the unswitched beam. The decay rate of the reflectivity could be fitted by a single exponential decay in all the three cases and was found to be the longest in Si (550\,ps) and the shortest in CdTe (75\,ps).

When output pulses below 50\,ps are desired, using CdTe as the first reflection switch provided the best background suppression ratio ($>$\,1000:1) thanks to its fast reflectivity decay. The background suppression was measured through cross-correlation of the double sliced pulse shown in Fig.~\ref{fig_5}b by averaging over a time window of 50\,ps on the pulse and 50\,ps after the trailing edge of the pulse. On the other hand, for output pulse durations in excess of 50\,ps, a Si reflection switch was preferred to maintain high reflectivity for a longer time. In this case, due to the large penetration depth of the near-infrared excitation pulse, thick plasma layers are formed so that transmission of the MIR radiation is negligible~\cite{rollandGeneration130fsecMidinfrared1986}. This makes Si a good candidate also for the second switch, both for the generation of short and long pulses. As this switch is used to suppress the trailing edge of the pulse, it is most effective if the radiation is to be blocked for as long as possible after photoexcitation.

The pulse lengths of double sliced pulses obtained with two silicon switches could be tuned over a wide range from $\sim$\,5\,ps to $\sim$\,1\,ns whilst for durations up to 300\,ps almost flat-top-intensity pulses were achieved (see time profiles in Fig.~\ref{fig_5}b). Although the minimum rise and fall time of the mid-infrared pulse is determined by the near infrared excitation pulse duration, in our setup the two pulses impinge on the semiconductors with a 13° non-collinear angle leading to a few picosecond temporal spread on the semiconductor surface.\\
The damage threshold was found to be significantly material dependent being just above 5\,mJ/cm² for Si, 3.9\,mJ/cm² for CdTe, and 1.7\,mJ/cm² for Ge. For fluence levels below damage threshold a maximum switching efficiency of 50\,\% was achieved for all materials.

\section*{MIR-pump THz-probe experiment on InSb}
\begin{figure}
\includegraphics[width=\linewidth]{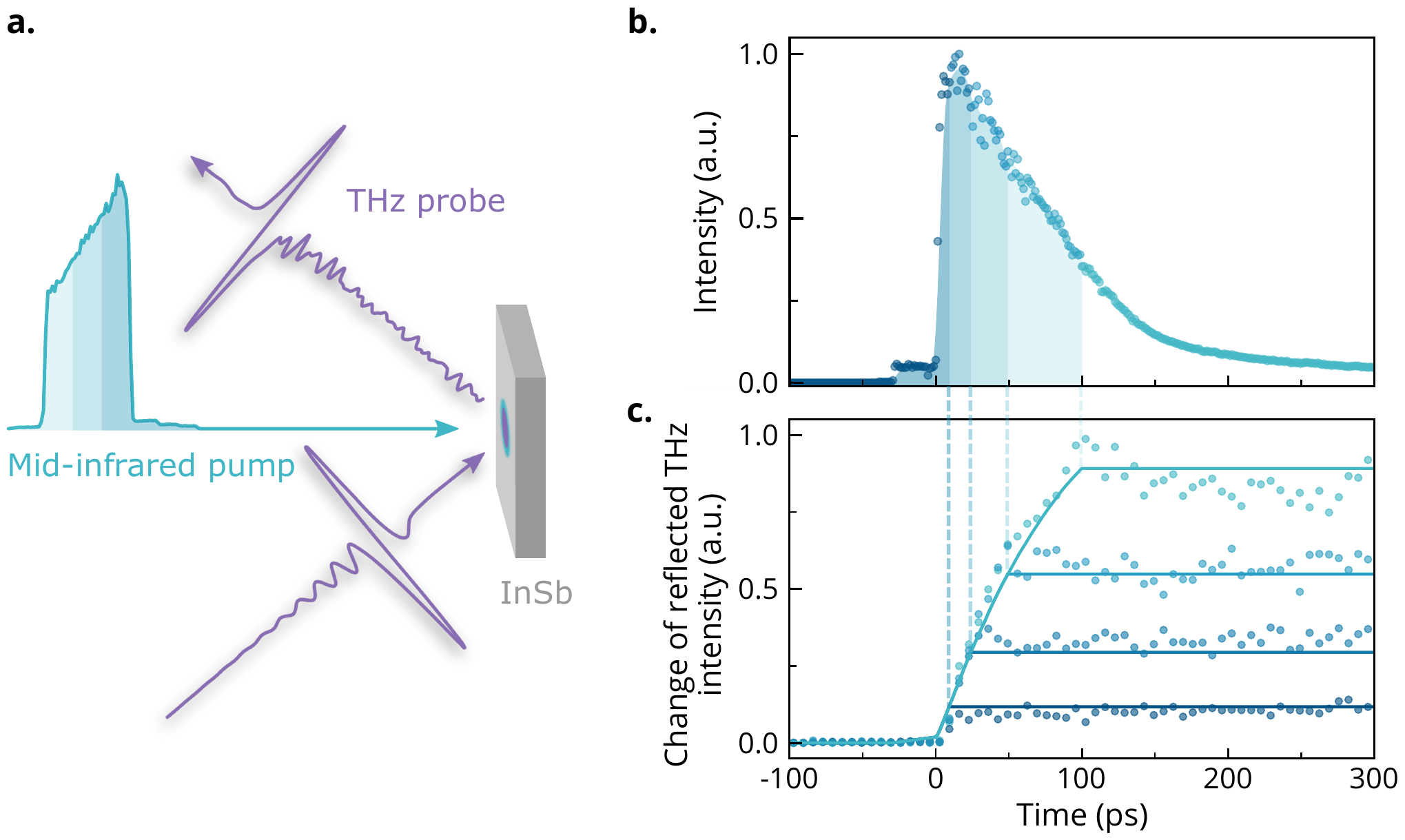}
\caption{\label{fig_6}Optical integrator on InSb
\textbf{a.} Schematic of the experimental setup for pulse length tunable mid-infrared pump, THz probe spectroscopy \textbf{b.} Cross-correlation measurement of a 10.6\,$\mathrm{\mu m}$ laser pulse with 1.3\,ns pulse length that was reflection switched with 100\,fs long 800\,nm wavelength pulses on CdTe. The decay of the semiconductor reflectivity due to carrier diffusion results in the declining slope of the reflected pulse. The different shadings indicate the pulse shapes that remain when using a second transmission switch after 10\,ps, 20\,ps, 50\,ps and 100\,ps. \textbf{c.} Measured change of reflected THz intensity of an InSb sample upon excitation with the pulses shown in a. (dots). Solid lines represent the calculated integral of the pulse intensity measurement from Fig.~\ref{fig_6}a up to the respective pulse duration of 10\,ps, 20\,ps, 50\,ps and 100\,ps. The reflectivity change in InSb scales with the incident number of photons, not with the electric field.}
\end{figure}

To demonstrate the applicability of the setup for pump-probe experiments with adjustable excitation pulse-lengths, we measured the time-resolved reflection of THz radiation from an InSb wafer excited with 10.6\,$\mathrm{\mu m}$ pulses with durations varying between 5\,ps and 100\,ps (see Fig.~\ref{fig_6}). These pulses were generated using a CdTe reflection and a Si transmission switch and their time-intensity profile is shown in Fig.~\ref{fig_6}b. THz pulses with bandwidth from $\sim$\,0.5\,THz to $\sim$\,3\,THz were generated in a commercial photoconductive antenna excited with 100\,fs long pulses at 800\,nm wavelength. These THz pulses were focused on the InSb sample, and their electric field was measured after reflection by means of electro-optic sampling in a 1\,mm thick (110)-cut ZnTe crystal.\\
Upon irradiation with the 10.6\,$\mathrm{\mu m}$ pump pulses,
the reflected THz intensity (the square of the THz electric field) increased linearly with the integral of the pump pulse intensity cross-correlation measurements, which we determined to be linear with the number of pump photons.
After the 10.6\,$\mathrm{\mu m}$ pump pulses hit the InSb wafer, the respective reflectivity levels stayed nearly constant for the observable measurement time of several hundred picoseconds. Therefore, the InSb response acts like an optical integrator of the pump photons.

In summary, we demonstrated a hybrid optical device that generates high-power mid-infrared pulses based on seeded CO\textsubscript{2} lasers, with tunable pulse durations achieved with two semiconductor-based plasma mirrors.  This source allows the extension of ultrafast experiments from femtosecond drives to pulse durations of hundreds of picoseconds. As the mid infrared pulses were naturally synchronized in time to a femtosecond laser, this device made pump-probe experiments with sub-picosecond resolution possible. 

\section*{Funding}
The research leading to these results received funding from the European Research Council under the European Union’s Seventh Framework Programme (FP7/2007-2013)/ERC Grant Agreement No.~319286 (QMAC). We acknowledge support from the Deutsche Forschungsgemeinschaft (DFG) via the Cluster of Excellence ‘The Hamburg Centre for Ultra-fast Imaging’ (EXC 1074 – project ID~194651731).

\section*{Acknowledgements}
We thank Michele Buzzi and Guido Meier for their help with manuscript preparation. Furthermore, we are grateful to Michael Volkmann for his technical assistance in the construction of the new optical apparatus presented in this work. We also acknowledge Toru Matsuyama, Boris Fiedler and Birger Höhling for their support with electronics and Jörg Harms for his help with graphics. Additionally, we thank Yannis Laplace for support in the early stage of the experimental design.

\section*{Disclosures}
The authors declare no conflicts of interest.

\nocite{*}

\bibliography{literature}

\end{document}